\documentclass[english,aps,prb,reprint,superscriptaddress,showpacs]{revtex4-1}
\usepackage[T1]{fontenc}
\usepackage[latin9]{inputenc}
\usepackage{color}
\usepackage{babel}
\usepackage{units}
\usepackage{amsmath}
\usepackage{amssymb}
\usepackage{graphicx}
\usepackage[unicode=true,
 bookmarks=true,bookmarksnumbered=false,bookmarksopen=false,
 breaklinks=false,pdfborder={0 0 0},backref=false,colorlinks=true]
 {hyperref}
\hypersetup{
 pdfauthor={V. Zaretskey},
 urlcolor=blue,citecolor=blue,bookmarksopen=true,bookmarksopenlevel=1,pdfstartview=FitH}
\usepackage{breakurl}
\usepackage{epstopdf}

\makeatletter

\DeclareRobustCommand{\greektext}{%
  \fontencoding{LGR}\selectfont\def\encodingdefault{LGR}}
\DeclareRobustCommand{\textgreek}[1]{\leavevmode{\greektext #1}}
\DeclareFontEncoding{LGR}{}{}
\DeclareTextSymbol{\~}{LGR}{126}
\newcommand{\lyxmathsym}[1]{\ifmmode\begingroup\def\b@ld{bold}
  \text{\ifx\math@version\b@ld\bfseries\fi#1}\endgroup\else#1\fi}

\providecommand{\tabularnewline}{\\}

 \@ifundefined{textcolor}{}
 {%
   \definecolor{BLACK}{gray}{0}
   \definecolor{WHITE}{gray}{1}
   \definecolor{RED}{rgb}{1,0,0}
   \definecolor{GREEN}{rgb}{0,1,0}
   \definecolor{BLUE}{rgb}{0,0,1}
   \definecolor{CYAN}{cmyk}{1,0,0,0}
   \definecolor{MAGENTA}{cmyk}{0,1,0,0}
   \definecolor{YELLOW}{cmyk}{0,0,1,0}
 }

\makeatother

\begin{document}

\title{Spectroscopy of a Cooper-Pair Box Coupled to a Two-Level System Via
Charge and Critical Current}

\date{\today}

\author{V. Zaretskey}

\author{B. Suri}

\author{S. Novikov}

\affiliation{Department of Physics, University of Maryland, College Park, Maryland,
20742}

\affiliation{Laboratory for Physical Sciences, College Park, Maryland, 20740}

\author{F. C. Wellstood}

\affiliation{Department of Physics, University of Maryland, College Park, Maryland,
20742}

\affiliation{Joint Quantum Institute and Center for Nanophysics and Advanced Materials,
College Park, Maryland, 20742}

\author{B. S. Palmer}

\affiliation{Laboratory for Physical Sciences, College Park, Maryland, 20740}
\begin{abstract}
We report on the quadrupling of the transition spectrum of an $\mbox{Al}/\mbox{AlO}_{x}/\mbox{Al}$
Cooper-pair box (CPB) charge qubit in the $\unit[4.0-7.3]{GHz}$ frequency
range. The qubit was coupled to a quasi-lumped element Al superconducting
resonator and measured at a temperature of $\unit[25]{mK}$. We obtained
good matches between the observed spectrum and the spectra calculated
from a model Hamiltonian containing two distinct low excitation energy
two-level systems (TLS) coupled to the CPB. In our model, each TLS
has a charge that tunnels between two sites in a local potential and
induces a change in the CPB critical current. By fitting the model
to the spectrum, we have extracted microscopic parameters of the fluctuators
including the well asymmetry, tunneling rate, and a surprisingly large
fractional change ($30-40\%$) in the critical current ($\unit[12]{nA}$).
This large change is consistent with a Josephson junction with a non-uniform
tunnel barrier containing a few dominant conduction channels and a
TLS that modulates one of them.
\end{abstract}


\maketitle

\section*{Introduction}

Dissipation and dephasing from two-level systems (TLS) are a serious
problem in many superconducting qubits. The aggregate effect of many
weakly coupled fluctuators causes $1/f$ charge noise, broadband dielectric
loss, and magnetic flux noise, as well as inhomogeneous broadening
and decreased measurement fidelity in qubits.\cite{ithier2005decoherence,simmonds2004decoherence,martinis2005decoherence,plourde2005fluxqubits,tian2007josephson,deppe2007phasecoherent}
An individual TLS quantum-coherently coupled to a qubit can typically
be identified when it leads to a resolvable avoided crossing in the
qubit spectrum. Such avoided level crossings have been observed in
phase,\cite{cooper2004observation,palomaki2010multilevel,simmonds2004decoherence,steffen2006statetomography}
flux,\cite{plourde2005fluxqubits,deppe2007phasecoherent} charge,\cite{kim2008anomalous}
quantronium,\cite{ithier2005decoherence} and transmon\cite{schreier2008suppressing}
qubits. While qubit performance is typically severely degraded near
such an avoided crossing,\cite{kim2008anomalous,sillanpaa2007coherent,simmonds2004decoherence,oh2006elimination,kline2009josephson}
strong qubit-TLS interactions allow the microscopic details of the
TLS to be determined.\cite{kim2008anomalous,shalibo2010lifetime,lupascu2009oneand}
Coherent coupling to a long-lived TLS also makes it possible to observe
coherent oscillations between a qubit and a TLS\cite{cooper2004observation}
or use the TLS as a quantum memory.\cite{neeley2008process,zagoskin2006quantum}

Two-level fluctuators in superconducting devices can be classified
into three types---charge, flux, or critical current---depending on
the nature of the interaction with the qubit. The microscopic origin
of charge and critical current fluctuators is believed to be impurity
ions such as H\cite{paik2010reducing,phillips1972tunneling} or low
coordination bonds in the amorphous dielectric used to build the devices.
In phase and flux qubits, it appears to be possible in principle but
difficult in practice to identify the exact nature of the qubit-TLS
interaction. In contrast, detailed spectroscopy on charge qubits or
Cooper-pair boxes (CPB) has enabled the identification of discrete
charge fluctuators.\cite{kim2008anomalous} For example, Kim \textit{et
al}. found TLS's that behaved as pure charge fluctuators.\cite{kim2008anomalous}
A moving charge could also modulate the critical current if it was
located in the tunnel barrier.\cite{constantin2007microscopic} Critical
current fluctuations have been frequently seen in Josephson junction
devices,\cite{clarke1996squidfundamentals,savo1987lowfrequency,wakai1986lowfrequency}
but apparently not in Josephson based qubits. This may be due to the
difficulty of conclusively distinguishing a critical current fluctuator
from a charge fluctuator. Alternatively, the relatively small area
of qubit junctions compared to that of conventional junctions leads
to far fewer total fluctuators and a corresponding decrease in the
probability of observing one. Also, qubit measurements are typically
made at less than $\unit[100]{mK}$, where critical current fluctuators
appear to be frozen out. Josephson junctions are a fundamental building
block of all superconducting qubits and an understanding of the origin
of critical current fluctuations is important for continued improvement
of qubit performance.

In this paper we report on a CPB with an unusual spectrum that has
multiple spectroscopic features displaced in both frequency and in
gate charge instead of an avoided level crossing. We find that the
spectrum, including the curvature of the spectral features, can be
modeled well with a critical current fluctuator coupled to a CPB with
an excitation energy for the fluctuator much less than the qubit energy.
By fitting our model to the spectrum we extract microscopic parameters
for the fluctuators.

\section*{Cooper-pair Box Qubit and Readout}

\begin{figure*}
\includegraphics[width=1\textwidth]{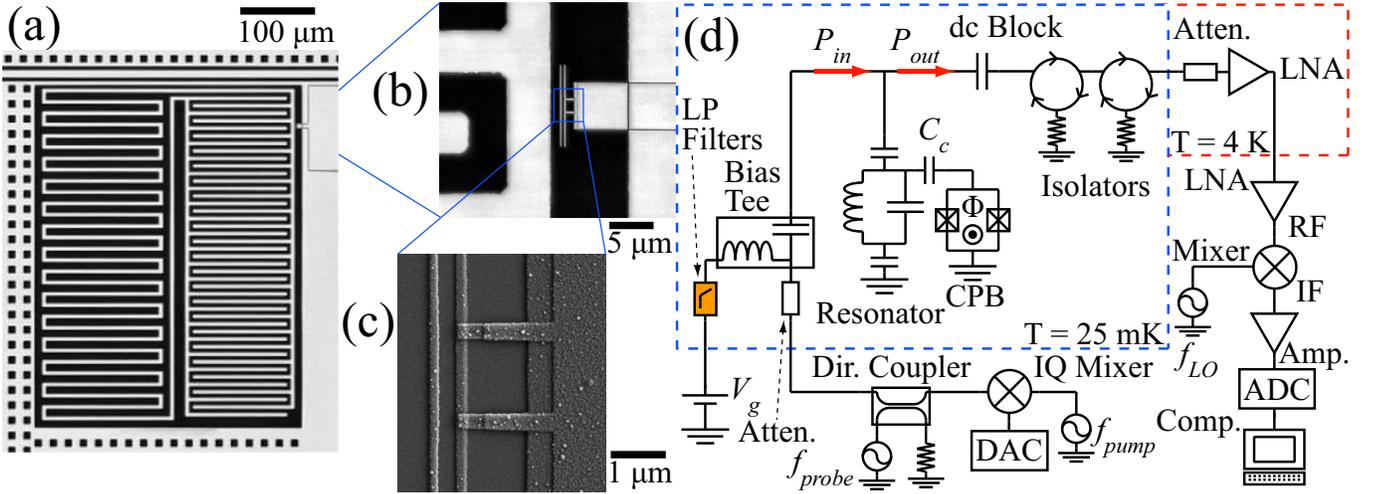}

\caption{(a) Optical image of the lumped element resonator coupled to a CPW
transmission line and surrounded by a perforated ground plane. Light
regions are aluminum metalization and dark are sapphire substrate.
(b) Optical image of the CPB located between the capacitor and ground
plane. (c) Scanning electron image of the CPB. The twinned features
are a consequence of the double-angle evaporation and the Josephson
junctions are located at the overlap of the two patterns. (d) Simplified
schematic of the experimental setup. The CPB is coupled through capacitor
$C_{c}$ to a quasi-lumped element LC resonator. Its state is readout
via a coherent heterodyne measurement of a microwave power at frequency
$f_{probe}$ transmitted through the device, amplified, mixed with
a local oscillator at frequency $f_{LO}$ and finally digitized. The
CPB transition frequency is controlled by the gate voltage $V_{g}$
and an external magnetic flux $\Phi$ and its state is coherently
manipulated using shaped microwave pulses at frequency $f_{pump}$.\label{fig:images+setup}}
\end{figure*}
Our Cooper-pair box (CPB) consists of a superconducting island connected
to a superconducting reservoir (ground) through two ultrasmall Josephson
tunnel junctions (critical current $I_{0}/2$ and junction capacitance
$C_{j}/2$) {[}see Fig. \ref{fig:images+setup}(d){]}. We can apply
gate voltage $V_{g}$ to a capacitively coupled gate (capacitance
$C_{g}$ to the island) to control the system's electrostatic energy.
Applying flux $\Phi$ to the loop formed by the two junctions tunes
the effective total critical current $I_{0}$ via the relation $I_{0}=I_{0}^{max}\cos\left(\pi\Phi/\Phi_{0}\right)$
where $\Phi_{0}=h/2e$ is the magnetic flux quantum.

Neglecting quasiparticle states, the Hamiltonian describing a CPB
in the charge basis is given by\cite{bouchiat1998quantum}
\begin{eqnarray}
\hat{H}_{CPB} & = & E_{c}\sum_{n}\left(2n-n_{g}\right)^{2}\left|n\right\rangle \left\langle n\right|\nonumber \\
 &  & -\frac{E_{J}}{2}\sum_{n}\left(\left|n+1\right\rangle \left\langle n\right|+\left|n\right\rangle \left\langle n+1\right|\right)\label{eq:hcpbcharge}
\end{eqnarray}
where $E_{c}=e^{2}/2C_{\Sigma}$ is the charging energy, $E_{J}=I_{0}\Phi_{0}/2\pi$
is the Josephson energy, $C_{\Sigma}=C_{j}+C_{g}$ is the total island
capacitance to ground, $n_{g}=C_{g}V_{g}/e$ is the reduced gate voltage
and $\left|n\right\rangle $ is the excess number of Cooper-pairs
on the island. For $E_{c}>E_{J}$ the system is highly anharmonic
and only a few charge states are needed to accurately describe the
lowest energy states. For charge qubits with $E_{c}\gg E_{J}$ and
$0.5<n_{g}<1.5$, the Hamiltonian can be reduced to\cite{bouchiat1998quantum}
\begin{equation}
\mathbb{H}_{CPB}=\begin{pmatrix}E_{c}\left(0-n_{g}\right)^{2} & -E_{J}/2\\
-E_{J}/2 & E_{c}\left(2-n_{g}\right)^{2}
\end{pmatrix}\label{eq:hcpbmatrx}
\end{equation}
which yields the excited state transition energy $\hbar\omega_{CPB}\left(n_{g}\right)=\sqrt{\left(4E_{c}\left(1-n_{g}\right)\right)^{2}+E_{J}^{2}}$.
Near the charge degeneracy point $n_{g}=1$ the transition energy
varies parabolically as $\hbar\omega_{CPB}\left(n_{g}\right)\approx E_{J}+8E_{c}^{2}\left(1-n_{g}\right)^{2}/E_{J}$.

To measure the state of the qubit, we coupled our qubit to a thin-film
quasi-lumped element LC resonator {[}see Fig. \ref{fig:images+setup}(d){]}
that was in turn weakly coupled to a microwave transmission line patterned
on the sample chip. To read out the state of the qubit, we apply microwave
power at the resonance frequency of the resonator and record the transmitted
microwave signal $S_{21}$.\cite{gambetta2007protocols} This is a
dispersive readout in which the qubit state modulates the resonance
frequency of the resonator. In the case of weak qubit-resonator coupling
$g$ and large detuning $\Delta=\omega_{CPB}-\omega_{r}$ the combined
CPB-resonator system Hamiltonian is approximately\cite{wallraff2005approaching,blais2004cavityquantum}
\begin{equation}
\hat{H}=\hbar\left(\omega_{r}+\frac{g^{2}}{\Delta}\sigma_{z}\right)\left(a^{\dagger}a+\frac{1}{2}\right)+\frac{\hbar\omega_{CPB}}{2}\sigma_{z}\label{eq:htotal}
\end{equation}
where $\hbar g=\left(2E_{c}C_{c}/e\right)\sqrt{\hbar\omega_{r}/2C}$
is the strength of the qubit-resonator coupling energy, $C_{c}$ is
the coupling capacitance between the resonator and the island of the
CPB, $C$ is the capacitance of the LC resonator, $\omega_{r}$ is
the resonance frequency, $a^{\dagger}a$ is the number operator for
excitations in the resonator, and $\sigma_{z}$ is the Pauli spin
operator. This Jaynes-Cummings Hamiltonian yields transitions in which
the bare resonator frequency $\omega_{r}$ is dispersively shifted
by $\chi=\pm g^{2}/\Delta$ depending on the state of the qubit. If
$\chi<\Gamma$, where $\Gamma$ is the resonator linewidth, the average
phase of the transmitted signal at $\omega=\omega_{r}$ is linearly
dependent on the excited state occupation probability. On the other
hand if $\chi>\Gamma$, then the in-phase or quadrature transmitted
voltage is proportional to the excited state occupation probability.\cite{gambetta2007protocols}

Additional complications can arise when the qubit and resonator are
coupled to another quantum system, such as a TLS. If multiple energy
levels in the combined system have similar detunings from the resonator,
the effective dispersive shift $\chi_{\mbox{eff}}$ will have a contribution
from each level.\cite{koch2007chargeinsensitive} Qubit state readout
can still be performed as described for the two level case, but the
sensitivity to a particular state depends on the choice of resonator
probe frequency. As we see below, this is our situation.

\section*{Charge and Critical Current TLS Model}

\begin{figure}
\includegraphics[width=1\columnwidth]{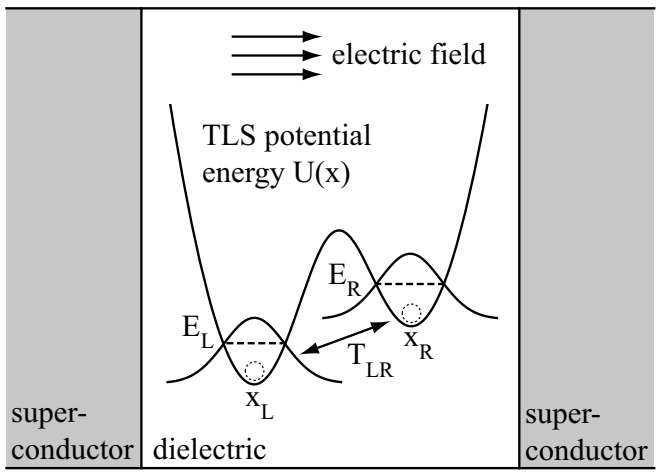}

\caption{Simplified diagram of the potential energy of a charged TLS in an
insulating tunnel barrier. The fluctuator can be localized at positions
$x_{R}$ or $x_{L}$ with corresponding energies $E_{R}$ or $E_{L}$
and can tunnel between them with energy $T_{LR}$. Additionally the
Josephson energy $E_{J}$ of the CPB depends on whether the fluctuator
is at $x_{R}$ or $x_{L}$.\label{fig:tls-diagram}}
\end{figure}
To include the effects on a CPB produced by a combined charge and
critical current fluctuator, we expand on the charge defect model
previously reported by Z. Kim, \textit{et al}.\cite{kim2008anomalous,wellstood2008spectroscopy}
We assume the fluctuator acts as a two-level system in which a point
charge in the tunnel barrier can tunnel between two potential well
minima. In the TLS position basis the fluctuator Hamiltonian is given
by
\begin{equation}
\mathbb{H}_{TLS}=\begin{pmatrix}E_{L} & T_{LR}\\
T_{LR} & E_{R}
\end{pmatrix}\label{eq:htlsmatrx}
\end{equation}
where $E_{L}$ and $E_{R}$ are energies of the charge in the left
and right position states and $T_{LR}$ is the tunneling matrix element
{[}see Fig. \ref{fig:tls-diagram}{]}. For an isolated fluctuator
the excited state transition energy is given by $\hbar\omega_{TLS}=\sqrt{\left(E_{R}-E_{L}\right)^{2}+4T_{LR}^{2}}$.

The charge coupling between the CPB and TLS originates from changes
in the electrostatic potential when the defect tunnels between its
two sites. Using Green's reciprocation theorem\cite{jackson1999classical}
the change in the induced polarization charge on the island of the
CPB when the fluctuator tunnels from the left to the right well is
$\Delta Q_{pi}=Q_{TLS}\left(x_{R}-x_{L}\right)\cos\left(\eta\right)/d$
where $Q_{TLS}$ is the TLS charge, $\eta$ is the angle the TLS displacement
vector makes relative to the electric field in the junction and $d\approx\unit[1]{nm}$
is the thickness of the tunnel junction. For fixed net charge on the
island this in turn results in a change in the electrostatic potential
of the island given by
\begin{equation}
\Delta V_{i}=\frac{Q_{TLS}}{C_{\Sigma}}\frac{\left(x_{R}-x_{L}\right)\cos\left(\eta\right)}{d}\mbox{.}\label{eq:dvi}
\end{equation}
Accounting for the electrostatic charging energy and the work done
by the gate voltage source when the point charge moves, the coupling
Hamiltonian is given by
\begin{equation}
\hat{H}_{CPB-TLS}=2E_{c}\left(2\hat{N}-n_{g}\right)\frac{Q_{TLS}}{e}\frac{\hat{x}\cos\left(\eta\right)}{d}\label{eq:hcoupl}
\end{equation}
where $\hat{N}$ is the CPB charge operator that counts the number
of excess Cooper-pairs on the island and $\hat{x}$ is the TLS position
operator.\cite{wellstood2008spectroscopy}

Combining Eqs. (\ref{eq:hcpbmatrx}), (\ref{eq:htlsmatrx}) and (\ref{eq:hcoupl})
we can write the total Hamiltonian for a CPB coupled to a single charge
fluctuator as $\hat{H}=\hat{H}_{CPB}+\hat{H}_{TLS}+\hat{H}_{CPB-TLS}$.
In block matrix form this becomes
\begin{equation}
\mathbb{H}=\begin{pmatrix}\mathbb{H}_{L} & \mathbb{T}\\
\mathbb{T} & \mathbb{H}_{R}
\end{pmatrix}\label{eq:hcpb1tls}
\end{equation}
where $\mathbb{T}=T_{LR}\mathbb{I}$, $\mathbb{I}$ is the $2\times2$
identity matrix, and $\mathbb{H}_{L}$ and $\mathbb{H}_{R}$ are the
CPB Hamiltonian with the TLS in either the left or right well. If
we assume $E_{L}=0$ then $\mathbb{H}_{L}=\mathbb{H}_{CPB}$ as given
by Eq. (\ref{eq:hcpbmatrx}) and
\begin{widetext}
\begin{equation}
\mathbb{H}_{R}=\begin{pmatrix}E_{c}\left(0-n_{g}\right)^{2}+E_{int}\left(0-n_{g}\right)+E_{R} & -E_{J}/2\\
-E_{J}/2 & E_{c}\left(2-n_{g}\right)^{2}+E_{int}\left(2-n_{g}\right)+E_{R}
\end{pmatrix}\label{eq:qrmatrx}
\end{equation}

\end{widetext}
\noindent where $E_{int}=2E_{c}Q_{TLS}\left(x_{R}-x_{L}\right)\cos\left(\eta\right)/ed$
sets the energy scale for the charge coupled interaction between the
fluctuator and the CPB.

If the TLS is in the junction tunnel barrier, it can also modulate
the critical current depending on its position.\cite{constantin2007microscopic}
This coupling can be accounted for by making the substitution $E_{J}\rightarrow E_{J}+\Delta E_{J}/2$
in $\mathbb{H}_{L}$ and $E_{J}\rightarrow E_{J}-\Delta E_{J}/2$
in $\mathbb{H}_{R}$.

Numerically diagonalizing the resulting $4\times4$ Hamiltonian $\mathbb{H}$,
we find the energy levels and the transition frequencies from the
ground state to the excited states of the system. An avoided crossing
occurs if the excited state of the TLS is resonant with the first
excited state of the CPB at some value of the gate voltage $n_{g}$.\cite{kim2008anomalous,wellstood2008spectroscopy}
However if the TLS excited state energy lies below the CPB transition
minimum the CPB spectrum is twinned, with one parabola corresponding
primarily to the excited state of the CPB and the other to a joint
excitation of the CPB and the TLS. Considered individually, each parabola
bears a strong resemblance to the spectrum of a TLS-free CPB. When
the tunneling energy $T_{LR}$ is small we can identify the qualitative
effects of each parameter on the twinned parabolas. $\Delta E_{J}$
creates an offset along the frequency axis and a change in the effective
curvature while $E_{int}$ creates an offset along the $n_{g}$ axis
and ``tilts'' the parabolas. $E_{R}$ also creates an offset along
the frequency axis that adds to or subtracts from the effect of $\Delta E_{J}$.
Finally $T_{LR}$ determines the size of any avoided crossings that
are present in the spectrum and determines the transition rate induced
by a gate perturbation between the ground state and excited states
involving the TLS.

We can further extend the model by considering the effect of two critical
current fluctuators. This is motivated by the observation of quadrupling
of the spectral lines in our data which can't be explained by the
presence of a single TLS. The total Hamiltonian for a CPB coupled
to two fluctuators in block matrix form is
\begin{equation}
\mathbb{H}=\begin{pmatrix}\mathbb{H}_{LL} & \mathbb{T}_{1} & \mathbb{T}_{2} & \mathbb{T}_{12}\\
\mathbb{T}_{1} & \mathbb{H}_{RL} & \mathbb{T}_{12} & \mathbb{T}_{2}\\
\mathbb{T}_{2} & \mathbb{T}_{12} & \mathbb{H}_{LR} & \mathbb{T}_{1}\\
\mathbb{T}_{12} & \mathbb{T}_{2} & \mathbb{T}_{1} & \mathbb{H}_{RR}
\end{pmatrix}\label{eq:hcpb2tls}
\end{equation}
where $\mathbb{T}_{1}=T_{LR,1}\mathbb{I}$, $\mathbb{T}_{2}=T_{LR,2}\mathbb{I}$,
and $\mathbb{T}_{12}=T_{12}\mathbb{I}$ where $T_{12}$ accounts for
any possible TLS-TLS coupling and the indices refer to the first or
second TLS. $\mathbb{H}_{ij}$ with $i,j\in\left\{ L,R\right\} $
is the CPB Hamiltonian with the respective TLS in either the left
or right well. For example, $\mathbb{H}_{RL}$ is given by
\begin{widetext}
\begin{equation}
\mathbb{H}_{RL}=\begin{pmatrix}E_{c}\left(0-n_{g}\right)^{2}+E_{int,1}\left(0-n_{g}\right)+E_{R,1} & -\left(E_{J}-\Delta E_{J,1}/2+\Delta E_{J,2}/2\right)/2\\
-\left(E_{J}-\Delta E_{J,1}/2+\Delta E_{J,2}/2\right)/2 & E_{c}\left(2-n_{g}\right)^{2}+E_{int,1}\left(2-n_{g}\right)+E_{R,1}
\end{pmatrix}\label{eq:q(rl)}
\end{equation}

\end{widetext}
\noindent and $\mathbb{H}_{LR}$ has the respective indices swapped.
$\mathbb{H}_{RR}$ includes the contribution of both TLS and in addition
present on the diagonal is a CPB mediated TLS-TLS interaction term\cite{wellstood2008spectroscopy}
of the form $E_{int,1}E_{int,2}/2E_{c}$.

\section*{Experimental Details}

We fabricated a thin-film lumped-element superconducting microwave
resonator using standard photolithography and lift-off techniques.
It was made from a $\unit[100]{nm}$ thick film of thermally evaporated
Al on a c-plane sapphire wafer that was patterned into a meander inductor
($L\approx\unit[2]{nH}$) and interdigital capacitor ($C\approx\unit[400]{fF}$)
coupled to a coplanar waveguide transmission line {[}see Fig. \ref{fig:images+setup}(a,b){]}.
The resonance frequency was $\omega_{r}/2\pi=\unit[5.47]{GHz}$ with
loaded quality factor $Q_{L}=35,000$, external quality factor $Q_{e}=47,000$,
and internal quality factor $Q_{i}=147,000$.

The CPB was subsequently defined by e-beam lithography and deposited
using double-angle evaporation and thermal oxidation of aluminum to
create the $\unit[350]{nm}\times\unit[150]{nm}$ Josephson tunnel
junctions {[}see Fig. \ref{fig:images+setup}(c){]}.\cite{dolan1977offsetmasks}
For the e-beam lithography we used a bilayer stack of MMA(8.5)MAA
copolymer and ZEP520A e-beam resist to facilitate lift-off and reduce
proximity exposure during writing. A $\unit[30]{nm}$ thick Al island
and $\unit[50]{nm}$ thick Al leads were deposited in an e-beam evaporator.
As discussed below, measurements of the CPB yielded $E_{c}/h$ in
the $\unit[4.4-5.3]{GHz}$ range and we tuned $E_{J}/h$ from $\unit[4]{GHz}$
to $E_{J}^{max}/h=\unit[7.33]{GHz}$.

The chip was enclosed in a rf-tight Cu box that was anchored to the
mixing chamber of an Oxford Instruments model 100 dilution refrigerator
at $\unit[25]{mK}$. Connections to the chip were made with Al wirebonds.
We used cold attenuators on the input microwave line and isolators
on the output line to filter thermal noise from higher temperatures
{[}see Fig. \ref{fig:images+setup}(d){]}. A filtered dc bias voltage
line was coupled to the input line using a bias tee before the device
and a dc block was placed after the sample box.

For spectroscopic measurements the resonator was probed with a weak
continuous microwave signal while a second pump tone was applied to
excite the qubit. The transmitted microwave signal at the probe frequency
was amplified with a HEMT amplifier%
\footnote{Weinreb (Caltech) Radiometer Group Low-Noise Amplifier. 2012. URL:
http://radiometer.caltech.edu/%
} sitting in the He bath {[}see Fig. \ref{fig:images+setup}(d){]}.
We implemented a coherent heterodyne setup to record the phase and
amplitude of the transmitted probe signal at $\unit[500]{ns}$ time
steps. After the HEMT, the signal was further amplified at room temperature,
mixed with a local oscillator tone to an intermediate frequency of
$\unit[2]{MHz}$ and then digitally sampled at a typical sampling
rate of $\unit[20]{MSa/s}$. A reference tone split off from the probe
signal was directly mixed and digitally sampled. Both signals passed
though a second stage of digital demodulation on a computer to extract
the amplitude and phase. All components were locked to a $\unit[10]{MHz}$
Rb atomic clock.%
\footnote{Stanford Research Systems (SRS) model FS725 Rubidium Frequency Standard%
} Both the probe and pump tone powers were optimized for ease of data
acquisition while also minimally disturbing the qubit. The probe tone
power was calibrated via the ac Stark shift.\cite{schuster2005acstark}
During measurement of the qubit state, the probe tone power was set
to populate the resonator with an average $\bar{n}=25$ photons while
the concurrent pump tone power was slightly above that needed to saturate
the CPB transitions.

\section*{Spectrum Characterization}

We measured the transition spectrum of the qubit by recording the
transmitted microwave probe signal while sweeping the dc gate voltage
and stepping the frequency of the second microwave pump signal. Fig.
\ref{fig:cpb+tls-fits}(d) shows a plot of the transmitted probe signal
amplitude as functions of $n_{g}$ and pump frequency $f_{pump}$
with $E_{J}$ tuned near $E_{J}^{max}$. Several unexpected and anomalous
features are evident. Rather than a single parabola, we observed two
parabolas with varying curvatures offset by $\approx\unit[0.25]{GHz}$
in frequency and $\approx0.04e$ in charge. This spectral structure
was stable over the course of four months and persisted as we tuned
the transition frequency from $\unit[4.0-7.3]{GHz}$. Close examination
of the figure reveals sections of two more quite weak parabolas. A
notable change in the spectrum occurred when we tuned $E_{J}$ to
bring the transition frequency below that of the resonator. As seen
in Fig. \ref{fig:cpb+tls-fits}(a), four parabolas are clearly visible
with the stronger new pair displaced $\approx\unit[0.40]{GHz}$ below
the original two. We note two additional anomalies we observed. First,
a ``dead zone'' was present between $\approx\unit[5.6-6.5]{GHz}$
where no spectrum was visible. Second, only half of the spectral parabolas---one
from each pair---were visible when measured with a pulsed probe readout.
For instance, in Fig. \ref{fig:cpb+tls-fits}(d) both parabolas were
present when we used a continuous measurement but only the bottom
parabola was visible when we used a pulsed measurement at a fixed
gate voltage $n_{g}=1$.%
\footnote{See online Supplemental Material for additional qubit spectrum plots.%
}

Some clues about the nature of the fluctuator are evident from an
examination of the spectrum. The frequency offset between the two
parabolas in Fig. \ref{fig:cpb+tls-fits}(d) could be caused by a
flux fluctuator that modulates $E_{J}$. However such a fluctuator's
effect on $E_{J}\left(\Phi\right)$ would be minimal when the applied
flux is near zero and increase as $E_{J}$ is reduced by an external
flux bias. As discussed below, this is the opposite of the behavior
we observed. Another argument against a simple flux fluctuator (such
as a vortex) or a simple charge fluctuator is that there are correlated
shifts in $n_{g}$ and frequency between the parabolas. In contrast,
the observed offsets and curvature changes are consistent with a two-level
system that is coupled to the CPB via both charge and critical current.

\begin{figure}
\includegraphics[width=1\columnwidth]{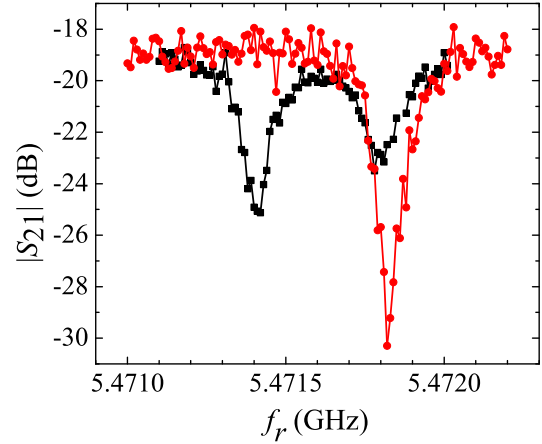}

\caption{Plot of the ratio of the transmitted output voltage to input voltage
($\left|S_{21}\right|$) versus frequency for two different preparations
of the qubit state. The filled black squares and black curve show
the transmission with the qubit biased at $n_{g}=1$ and driven to
a mixed ground and excited state. The two dips are resonances at $\omega_{r}^{\prime}\pm\chi_{\mbox{eff}}$.
The filled red circles and red curve show the transmission with the
qubit in the ground state and far detuned from the resonator at $n_{g}=0$.
The single dip is the bare resonator frequency at $\omega_{r}$ .\label{fig:ng-cuts}}
\end{figure}
Several factors indicate that the fluctuator is coherently coupled
to the CPB. An incoherently coupled low frequency critical current
fluctuator would be expected to produce twinning in the resonator
dispersive shift $\chi$ in addition to twinning of the spectral lines.
This twinning of the dispersive shift would be manifest either as
twinning of the ground state resonator frequency or broadening of
the resonator linewidth. We didn't observe either of these effects.
Instead we observed an effective dispersive shift $\chi_{\text{eff}}$
consistent with contributions from multiple levels {[}see Fig. \ref{fig:ng-cuts}{]}.
We determined the effective dispersive shift $\chi_{\text{eff}}$
and effective resonator frequency $\omega_{r}^{\prime}$ by recording
the resonator response with the qubit in the ground and excited states.
We also measured the bare resonator frequency $\omega_{r}$ by far
detuning the qubit from the resonator by biasing at $n_{g}=0$ {[}see
Fig. \ref{fig:ng-cuts}{]}. As expected $\omega_{r}^{\prime}\neq\omega_{r}$
and the effective dispersive shift $\chi_{\text{eff}}$ differed between
the excited states corresponding to the various parabolas. Finally,
in previous cases of incoherent fluctuator coupling we found that
the qubit was rendered inoperable.\cite{schuster2007circuit,kim2010dissipative}
Yet in this case we were able to measure qubit excited state lifetimes
$T_{1}$ in the $\unit[15-30]{\lyxmathsym{\textgreek{m}}s}$ range
and record Rabi oscillations for all of the parabolas.

The strength of the qubit-TLS coupling indicates that the TLS was
located close to the CPB Josephson junctions, either in the tunnel
barrier itself or on the surface of the CPB island. Furthermore, we
note that the spectra were $2e$ periodic in $n_{g}$. This is the
expected periodicity for a charge fluctuator that is in the tunnel
barrier\cite{wellstood2008spectroscopy}, and such a fluctuator would
need to be in the $\mbox{AlO}_{x}$ tunnel barrier to produce a critical
current change.

\section*{Fitting and Discussion}

\begin{figure*}
\includegraphics[width=1\textwidth]{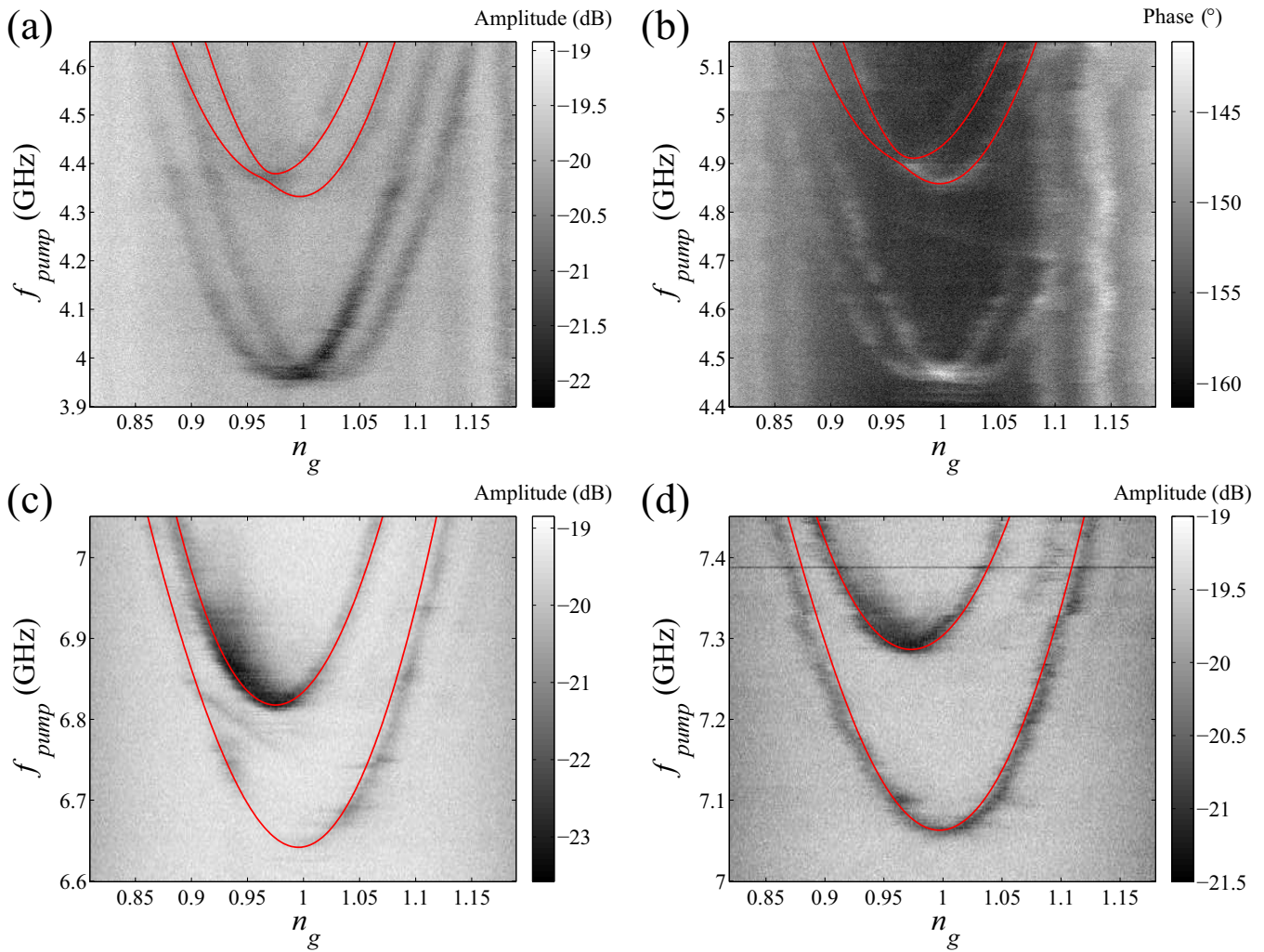}

\caption{Measured transition spectrum of the CPB at four different external
magnetic flux $\Phi$ bias values. The red lines are the theoretical
spectrum using a Hamiltonian consisting of a single charged two-level
fluctuator coupled to a CPB. In (a) and (b) (data sets \#1 and \#2)
the qubit transition frequency is tuned below the resonator and only
the top two parabolas were considered when fitting to the model. In
(c) and (d) (data sets \#3 and \#4) the qubit transition frequency
is tuned above the resonator. Table \ref{tab:cpb+1tls-fits} lists
the parameters used to generate each curve. (d) The dark horizontal
line at $\unit[7.39]{GHz}$ is a charge noise artifact.\label{fig:cpb+tls-fits}}

\end{figure*}
\begin{table}
\caption{Fit parameters for the model of a single two-level fluctuator coupled
to a CPB. The corresponding spectra are plotted in Fig. \ref{fig:cpb+tls-fits}.
$E_{c}$ and $E_{J}$ are the CPB charging and Josephson energies.
$E_{R}$ is the TLS potential energy well asymmetry, $E_{int}$ is
the charge coupling between the TLS and the CPB, $\Delta E_{J}$ is
the change in the CPB Josephson energy when the TLS tunnels between
wells and $T_{LR}$ is the TLS tunneling rate.\label{tab:cpb+1tls-fits}}

\centering{}%
\begin{tabular}{lllllllll}
\hline 
Data set &  & \#1 &  & \#2 &  & \#3 &  & \#4\tabularnewline
\hline 
$E_{c}/h$ (GHz) & \enskip{} & $4.5$ & \enskip{} & $4.5$ & \enskip{} & $4.5$ & \enskip{} & $4.5$\tabularnewline
$E_{J}/h$ (GHz) &  & $3.64$ &  & $4.16$ &  & $5.93$ &  & $6.33$\tabularnewline
$\Delta E_{J}/h$ (GHz) &  & $1.50$ &  & $1.54$ &  & $1.84$ &  & $2.02$\tabularnewline
$E_{R}/h$ (GHz) &  & $0.62$ &  & $0.62$ &  & $0.62$ &  & $0.62$\tabularnewline
$E_{int}/h$ (GHz) &  & $0.35$ &  & $0.35$ &  & $0.35$ &  & $0.35$\tabularnewline
$T_{LR}/h$ (GHz) &  & $0.01$ &  & $0.01$ &  & $0.06$ &  & $0.06$\tabularnewline
\hline 
\end{tabular}
\end{table}
We first fit the single TLS model to the measured spectrum at several
different external flux bias values. In our device $E_{c}$ is comparable
to $E_{J}$, so we needed to include $4$ charge states in the CPB
Hamiltonian block matrices to better approximate the CPB behavior.
We initially focused only on the top two parabolas to better understand
the effects of the model parameters and the relation between the one
and two TLS models. The solid red curves in Fig. \ref{fig:cpb+tls-fits}
show the predicted spectrum for those parabolas and the fits look
reasonable.

The optimal fit parameters are summarized in Table \ref{tab:cpb+1tls-fits}
and give reasonable results for all values of the flux bias. Individual
fit parameters could typically be varied by approximately 20\% while
maintaining a reasonable looking fit. The large uncertainty is partly
due to the fact that the frequency offset between the twinned parabolas
arises from both $\Delta E_{J}$ and $E_{R}$. Additionally the model
predicts avoided crossings which were too small to resolve, and this
meant we could place an upper bound on the TLS tunneling strength
$T_{LR}$. We note that the data sets with different applied flux
only require $E_{J}$ and $\Delta E_{J}$ to be adjusted, which is
consistent with changing flux bias, except for a change in $T_{LR}$
when the qubit is tuned from below to above the resonator $\omega_{r}$.
The model also predicts a nearly flat TLS spectral line in the $\unit[1-2]{GHz}$
range, roughly equal to the transition frequency of the isolated fluctuator.
We didn't observe such a feature, perhaps because our resonator perturbative
measurement technique was insensitive to a low frequency TLS-only
transition.

It is important to consider if other models can explain our observations.
We can eliminate a coherently coupled flux fluctuator using the same
reasoning used to exclude the incoherently coupled case. In particular
this suggests that the unusual spectrum isn't due to coupling to a
moving vortex. Another possibility is that the data could be fit by
a charged fluctuator with $\Delta E_{J}=0$. Such model would predict
a large ``tilt'' of the parabolas that disagrees with data covering
a wider $n_{g}$ and frequency span.

\begin{figure*}
\includegraphics[width=1\textwidth]{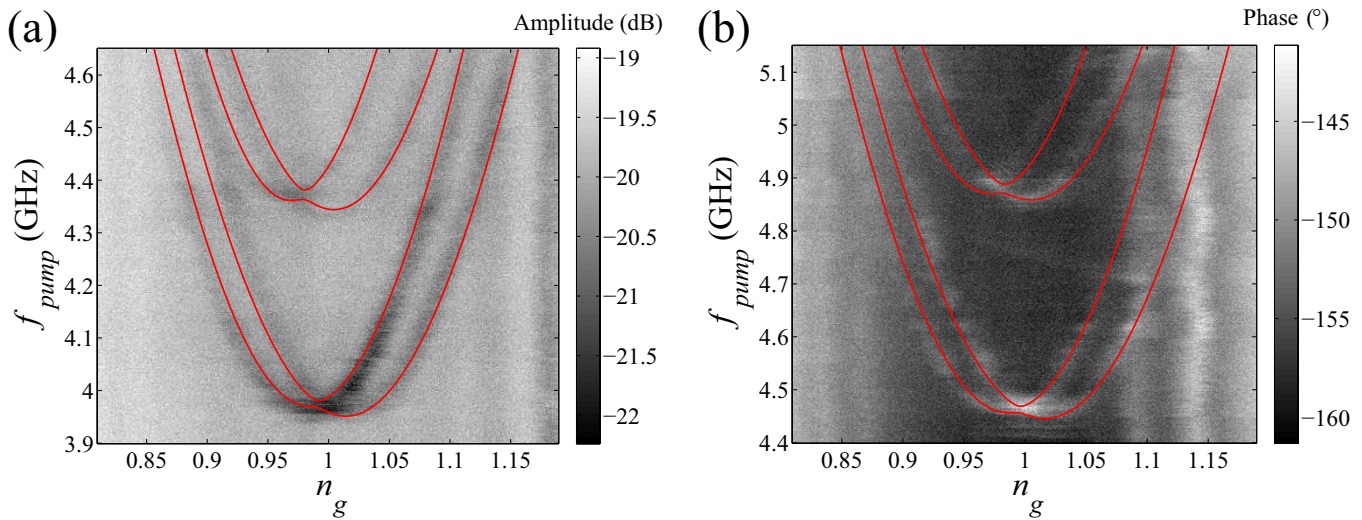}

\caption{Measured transition spectrum of the CPB at two different external
magnetic flux $\Phi$ bias values. The red lines are the theoretical
spectrum using a Hamiltonian consisting of two charged two-level fluctuators
coupled to a CPB. Plots (a) and (b) (data sets \#1 and \#2) are the
same data sets as plots (a) and (b) in Fig. \ref{fig:cpb+tls-fits}.
Table \ref{tab:cpb+2tls-fits} lists the parameters used to generate
each curve.\label{fig:cpb+2tls-fits}}
\end{figure*}
\begin{table}
\caption{Fit parameters for the model of two two-level fluctuators coupled
to a CPB. The corresponding spectra are plotted in Fig. \ref{fig:cpb+2tls-fits}.
$E_{c}$ and $E_{J}$ are the CPB charging and Josephson energies.
$E_{R,1}$ and $E_{R,2}$ are the potential energy well asymmetries
for the first and second TLS. Similarly $E_{int,1}$ and $E_{int,2}$
are the charge coupling between the first and the second TLS and the
CPB and $\Delta E_{J,1}$ and $\Delta E_{J,2}$ are the changes in
the CPB Josephson energy when the TLS tunnel between their respective
wells. $T_{LR,1}$ and $T_{LR,2}$ are the TLS tunneling rates while
$T_{12}$ is the TLS-TLS coupling strength.\label{tab:cpb+2tls-fits}}

\centering{}%
\begin{tabular}{lllll}
\hline 
Data set &  & \#1 &  & \#2\tabularnewline
\hline 
$E_{c}/h$ (GHz) & \enskip{} & $4.3$ & \enskip{} & $4.3$\tabularnewline
$E_{J}/h$ (GHz) &  & $2.79$ &  & $3.43$\tabularnewline
$\Delta E_{J,1}/h$ (GHz) &  & $1.36$ &  & $1.40$\tabularnewline
$E_{R,1}/h$ (GHz) &  & $0.62$ &  & $0.62$\tabularnewline
$E_{int,1}/h$ (GHz) &  & $-0.40$ &  & $-0.40$\tabularnewline
$T_{LR,1}/h$ (GHz) &  & $0.00$ &  & $0.00$\tabularnewline
$\Delta E_{J,2}/h$ (GHz) &  & $-1.00$ &  & $-0.68$\tabularnewline
$E_{R,2}/h$ (GHz) &  & $-0.82$ &  & $-0.69$\tabularnewline
$E_{int,2}/h$ (GHz) &  & $0.13$ &  & $0.15$\tabularnewline
$T_{LR,2}/h$ (GHz) &  & $0.04$ &  & $0.04$\tabularnewline
$T_{12}/h$ (GHz) &  & $0.04$ &  & $0.04$\tabularnewline
\hline 
\end{tabular}
\end{table}
We also fit the entire spectrum of four parabolas to the two TLS model
{[}see Eqs. \ref{eq:hcpb2tls} and \ref{eq:q(rl)}{]}. The solid red
curves in Fig. \ref{fig:cpb+2tls-fits} show the best fit spectrum
superposed on the data. The optimal fit parameters are summarized
in Table \ref{tab:cpb+2tls-fits}. The vertical lines at $n_{g}\approx1\pm0.09$
and $n_{g}\approx1\pm0.14$ are due to the resonant crossing between
the qubit parabolas and the resonator line at $\omega_{r}$. Although
the fits are reasonable and capture all of the major features, the
fit parameters contain one surprise. If we assume two independent
fluctuators, the simplest assumption in light of the strong shielding
of electric fields in the dielectric of the Josephson junction by
the superconducting electrodes, then we would expect $T_{LR,1}\neq0$
and $T_{LR,2}\neq0$ while $T_{12}=0$. However our fit yields $T_{LR,1}=0$
while $T_{LR,2}\neq0$ and $T_{12}\neq0$ which suggests coupled TLS's
or more complicated microscopic behavior. Furthermore, we note that
several of the TLS parameters, such as the charge coupling $E_{int,1}$,
change values when switching from the single to the double TLS model.
This indicates that the two TLS model is needed to explain the full
quadrupled spectrum and suggests that there is significant interaction
between the TLS\textquoteright{}s.

There are some noteworthy implications from the magnitude of the fit
parameters. First, we note that $\Delta E_{J}/E_{J}\approx30\%$.
The large relative size of $\Delta E_{J}$ to $E_{J}$ suggests that
the junction tunnel barrier is non-uniform with a few dominant conduction
channels and that the TLS is located near and modulates one of these
channels.\cite{dorneles2003theuse} Second, the TLS tunneling matrix
element $T_{LR}\lesssim\unit[0.06]{GHz}$ is small compared to the
other energies in the system, indicating that the TLS is tunneling
between fairly well isolated sites. We can also place a lower bound
on $T_{LR}$ by noting that for $T_{LR}<\unit[0.01]{GHz}$ the spectra
would be too faint to observe. If the excited state of such a TLS
were resonant with the first excited state of the CPB, the resulting
avoided crossing would be very small and difficult to resolve. Our
extracted tunneling matrix element values are also significantly smaller
than those reported by Z. Kim, \textit{et al}.,\cite{kim2008anomalous}
which were in the $\unit[3-13]{GHz}$ range. There is a similar relation
between the range of well asymmetry values extracted by us, $E_{R}=\unit[0.6-0.8]{GHz}$,
and those reported by Z. Kim, \textit{et al}., $E_{R}=\unit[7-39]{GHz}$.
Assuming a TLS charge of $Q_{TLS}=e$ and a tunnel barrier thickness
of $d=\unit[1]{nm}$, we estimate the maximum hopping distance of
the defect at $0.2-0.45$\,\AA. This is in agreement with the bounds
of $0.32-0.83$\,\AA\ found by Z. Kim, \textit{et al}.

Discrete critical current fluctuators have been reported in current
biased Josephson junctions, identified via either a random telegraph
signal in the voltage time trace or a signature Lorentzian bump in
the noise spectrum.\cite{eroms2006lowfrequency,gustafsson2011noiseproperties,savo1987lowfrequency,wakai1986lowfrequency}
One way we can compare our TLS's to others is to calculate the effective
defect area $A_{\text{eff}}$ given by $A_{\text{eff}}=\left(\Delta E_{J}/E_{J}\right)A_{j}$.
For our device find $A_{\text{eff}}\approx\unit[18,000]{nm^{2}}$
where $A_{j}=\unit[350\times150]{nm^{2}}$ is the junction area. This
value is much larger than the $A_{\text{eff}}=\unit[1-2]{nm^{2}}$
reported in similar junctions,\cite{eroms2006lowfrequency} the $A_{\text{eff}}\unit[\approx600]{nm^{2}}$
seen in larger area junctions,\cite{savo1987lowfrequency} or the
$A_{\text{eff}}=\unit[72]{nm^{2}}$ found in similar area $\mbox{high-}T_{c}$
superconductor grain boundary junctions.\cite{gustafsson2011noiseproperties}
On the other hand, the absolute value of the critical current fluctuation
$\Delta I_{0}\approx\unit[4]{nA}$ we observed is close to that reported
in both similar area ($\Delta I_{0}=\unit[9.2]{nA}$)\cite{wakai1986lowfrequency}
and larger junctions ($\Delta I_{0}\approx\unit[1]{nA}$).\cite{savo1987lowfrequency}
One notable difference that might account for some of these discrepancies
is that the critical current density of our sample ($\unit[23]{A/cm^{2}}$)
is smaller by an order of magnitude or more than the referenced samples.
If we assume that the conductance of a tunneling channel is similar
between the various devices, this is consistent with a small number
of tunneling hot spots in our junction.

\section*{Comments and Conclusion}

The longitudinal relaxation rate of a TLS in an amorphous solid is
expected to be limited by $1/T_{1}=\alpha\hbar\omega_{TLS}T_{LR}^{2}\coth\left(\hbar\omega_{TLS}/2k_{B}T\right)$
where $T$ is temperature and $\alpha$ is a material dependent constant.\cite{phillips1981amorphous}
From the results of Z. Kim, \textit{et al}.\cite{kim2008anomalous}
we estimate $1/\alpha\approx\unit[10^{2}]{\text{\textgreek{m}}s\cdot GHz^{3}}\cdot h^{3}$
for the dielectric $\mbox{AlO}_{x}$ in the tunnel junction barrier.
Our fit values then place an upper bound on the TLS excited state
lifetime of $T_{1}\lesssim\unit[1]{ms}$. This bound is consistent
with a relatively long TLS lifetime and with our qubit $T_{1}\approx\unit[15-30]{\text{\textgreek{m}}s}$.
The excited states of the system are mixtures of pure CPB and TLS
excited states, so the decay rate is a weighted average of the pure
CPB and TLS decay rates. For example, according to our fits to the
model at $n_{g}=1$ the lower parabola in Fig. \ref{fig:cpb+tls-fits}(d)
is composed of a $12\%$ CPB excitation and an $88\%$ joint CPB plus
TLS excitation while the upper parabola is an $88\%$ CPB excitation
and a $12\%$ joint CPB plus TLS excitation. Only when both the qubit
and TLS decay rates are small, as is our case, will the system decay
time be long in both parabolas.

Finally it's worthwhile to speculate why this behavior was observed
in our sample.%
\footnote{D. I. Schuster has observed similar spectral features in a CPB, suggesting
that this type of defect is rare, but not unique. (personal communication,
March 2012)%
} In order to observe spectral twinning rather than an avoided crossing,
the TLS needs to be coupled to the qubit but have a transition frequency
less than $E_{J}/h$. That this occurred is a statistical coincidence.
Observing two such defects in the same sample is less likely, and
the TLS fit parameters suggest they are correlated. Furthermore, we
are biased in selecting samples for detailed study that have especially
conspicuous features, such as large avoided crossings or anomalous
spectra, and the parameter values of such samples are likely to be
somewhat unusual.

While our simple model provides a good fit to the recorded spectrum,
it leaves other questions unanswered. The resonator wasn't included
in the model but some of our observations suggest that it may produce
significant effects on the spectrum. Inclusion of the resonator in
the model would allow a theoretical calculation of the expected dispersive
shift and a comparison with the data. A more complete model may also
elucidate the role, if any, the resonator played in the the large
difference in the visibility of the different parabolas when the qubit
was tuned from below to above the resonator $\omega_{r}$ or the ``dead
zone'' we observed between $\approx\unit[5.6-6.5]{GHz}$ where no
spectrum was visible. Perhaps the most puzzling feature was that half
of the spectral parabolas weren't visible when measured with a pulsed
probe. Unfortunately additional data on this issue wasn't obtained.

In conclusion we have examined the transition spectrum of a CPB that
had an anomalous quadrupling of the spectral lines. A microscopic
model of one or two charged critical current fluctuators coupled to
a CPB was used to fit the spectrum. The fits were in good agreement
with the data, reproduced the key features in the spectrum, and allowed
us to extract microscopic parameters for the TLS's. Our tunneling
terms were much smaller than those reported by Z. Kim, \textit{et
al}.\cite{kim2008anomalous} in their measurements of avoided crossings.
Finally, the large fractional change $\Delta E_{J}/E_{J}$ of $30-40\%$
suggests that the tunnel barrier is non-uniform in thickness with
the TLS hopping blocking a dominant conduction channel.
\begin{acknowledgments}
FCW would like to acknowledge support from the Joint Quantum Institute
and the State of Maryland through the Center for Nanophysics and Advanced
Materials. The authors would like to thank M. Khalil, Z. Kim, P. Nagornykh,
K. Osborn, and N. Siwak for many useful discussions.
\end{acknowledgments}
\bibliographystyle{apsrev4-1}

\end{document}